\documentstyle[epsfig]{elsart}

\begin{document}
\begin{frontmatter}

\title{Production of $\eta\prime$ Mesons in the $pp \rightarrow pp\eta\prime$ Reaction at 3.67 GeV/c} 

\author[Tor]{F.~Balestra}
\author[LNS]{Y.~Bedfer}
\author[LNS,Tor]{R.~Bertini}
\author[IUCF]{L.C.~Bland}
\author[Gie]{A.~Brenschede\thanksref{aa}}
             \thanks[aa]{current address: Brokat Infosystems AG - Stuttgart}
\author[LNS]{F.~Brochard\thanksref{ab}}
             \thanks[ab]{current address: LPNHE X, Ecole
                                          Polytechnique Palaiseau}
\author[Tor]{M.P.~Bussa}
\author[IUCF]{Seonho~Choi\thanksref{bb}}
            \thanks[bb]{current address: Temple University, Philadelphia}
\author[Kr1]{M.~Debowski\thanksref{bc}}
            \thanks[bc]{current address: FZ-Rossendorf}
\author[IUCF]{M.~Dzemidzic\thanksref{cc}}
            \thanks[cc]{current address: IU School of Medicine -
Indianapolis}
\author[LNS]{J.-Cl.~Faivre}
\author[JINR]{I.V.~Falomkin\thanksref{aaa}}
            \thanks[aaa]{deceased}
\author[UPO]{L.~Fava}
\author[Tor]{L.~Ferrero}
\author[Kr2,Kr1]{J.~Foryciarz\thanksref{dd}}
      \thanks[dd]{current address: Motorola Polska Software Center - Krak\'ow}
\author[Gie]{I.~Fr\"ohlich}
\author[JINR]{V.~Frolov}
\author[Tor]{R.~Garfagnini}
\author[Tor]{A.~Grasso}
\author[LNS]{S.~Heinz}
\author[JINR]{V.V.~Ivanov}
\author[IUCF]{W.W.~Jacobs}
\author[Gie]{W.~K\"uhn}
\author[Tor]{A.~Maggiora}
\author[Tor]{M.~Maggiora}
\author[LNS,Tor]{A.~Manara}
\author[UPO]{D.~Panzieri}
\author[Gie]{H.-W.~Pfaff}
\author[Tor]{G.~Piragino}
\author[JINR]{G.B.~Pontecorvo}
\author[JINR]{A.~Popov}
\author[Gie]{J.~Ritman}
\author[Kr1]{P.~Salabura}
\author[JINR]{V.~Tchalyshev}
\author[Tor]{F.~Tosello} 
\author[IUCF]{S.E.~Vigdor}
\author[Tor]{ G.~Zosi}

\address[JINR]{JINR, Dubna, Russia }
\address[IUCF]{Indiana University Cyclotron Facility, Bloomington, Indiana, U.S.A.}
\address[LNS]{Laboratoire National Saturne, CEA Saclay, France}
\address[Tor]{Dipartimento di Fisica "A. Avogadro" and INFN - Torino, Italy}
\address[UPO]{Universita' del Piemonte Orientale and INFN - Torino, Italy}
\address[Kr1]{M. Smoluchowski Institute of Physics, Jagellonian University, Krak\'ow, Poland}
\address[Kr2]{H.Niewodniczanski Institute of Nuclear Physics, Krak\'ow, Poland}
\address[Gie]{II. Physikalisches Institut,  University of Gie{\ss}en, Germany}
\address[TRI]{TRIUMF - Vancouver, Canada}

\collaboration{DISTO Collaboration}

\begin{abstract}
  The ratio of the total exclusive production cross sections for $\eta\prime$ 
  and $\eta$ mesons has been measured in the $pp$ reaction at
  $p_{beam}=3.67$~GeV/c.  The observed $\eta\prime/\eta$ ratio is
  $(0.83\pm{0.11}^{+0.23}_{-0.18})\times 10^{-2}$ from which the exclusive 
  $\eta\prime$ meson production cross section is determined to be 
  $(1.12\pm{0.15}^{+0.42}_{-0.31})\mu b$.
  Differential cross section distributions have been measured.
  Their shape is consistent with isotropic $\eta\prime$ meson production.

\end{abstract}

\begin{keyword}
 $\eta\prime$ meson, proton-proton final state interaction 
\PACS{14.40.Cs, 13.75.Cs, 25.40.Ve }
\end{keyword}

\end{frontmatter}

The study of the $\eta\prime$ meson  production is of particular interest
because of its large mass compared to the other members of the
ground state pseudoscalar meson nonet. The spontaneous breaking of
chiral symmetry causes the existence of massless Goldstone bosons, 
which acquire mass due to explicit chiral symmetry breaking, and are associated
with the pseudoscalar meson nonet.
In addition quantization effects in QCD lead to the so-called $U_{A}(1)$ anomaly,
which allows the $\eta\prime$ meson to gain mass by a different mechanism than the
Goldstone bosons~\cite{WEI,WIT,VEN,HOF,GLU}.
Nevertheless, the origin of the $\eta\prime$ mass and its structure in terms
of quark and gluon degrees of freedom remain controversial.

Recent measurements 
of the $\eta\prime$ meson by 
the CLEO collaboration show an anomalously large branching
ratio of B-mesons to $\eta\prime X$ and $\eta\prime K$ \cite{CLEO-D},
which might indicate a strong coupling of the $\eta\prime$ 
meson to gluons \cite{CLEO-T}.
Furthermore, the quark component of the nucleon's axial-vector
matrix element measured in the EMC experiment~\cite{EMC} suggests that 
the $\eta\prime$ meson couples very weakly to the nucleon ~\cite{HAT,SHO}.

First measurements ~\cite{MOS,HIB,MOS2} of the reaction 
$pp \rightarrow pp \eta\prime $ near the production threshold provide 
the possibility to determine the coupling constant $g_{\eta\prime NN}$. 
However, a 
quantitative evaluation of this coupling constant 
requires answers to several open questions concerning the
production mechanism,
such as the roles of
(i) meson-exchange currents, 
(ii) baryon resonances in the 
production mechanism (comparable with the role of the $N^{\star}(1535)$ for
$\eta$ meson production ~\cite{VET}), and (iii)
final-state interactions (FSI).

The existing data close to threshold are consistent with different 
one-boson-exchange models (OBE) including FSI
~\cite{MEI,SIB} given the ambiguities
in the treatment of heavy-meson-exchange currents.
Evaluation of the different models 
requires a consistent description over 
a wide energy range, but data have been lacking
at higher energies, where identification of the $\eta\prime$ production can
no longer be made solely by detecting two protons in a small forward
cone~\cite{MOS,HIB,MOS2}.

In this letter, we report on a measurement of 
the $\eta\prime/\eta$ production cross section 
ratio at an energy where proton-proton ($pp$) FSI have a much smaller
relative influence on the production mechanism~\cite{SIB,ETAFSI} compared to 
the near to threshold data~\cite{MOS,HIB,MOS2}. In addition, we show 
the differential cross sections of the $\eta\prime$ meson as a function of the
polar angle in the CM (center of mass) reference frame and the momentum
distributions of the final state particles. These distributions are related
to the partial wave contribution in the exit channel ($pp\eta\prime$).

We studied the $pp$ reaction at the SATURNE II accelerator facility at Saclay.
The proton beam of momentum $3.67$ GeV/c was incident on 
a liquid hydrogen target and charged products
were detected using the DISTO spectrometer, which is described in
detail elsewhere~\cite{NIM}. This spectrometer consisted of a large dipole
magnet ($40$ cm gap size, operating at $1.0$ Tm) which covered the target area and two
sets of scintillating fiber hodoscopes. Outside the magnetic field, two sets
of multi-wire proportional chambers (MWPC) were mounted, along
with segmented plastic scintillator hodoscopes 
and water \v Cerenkov detectors. 
The scintillator hodoscopes and the \v Cerenkov detectors allow particle
identification by combining either the energy loss, time of flight or \v Cerenkov light
output with the particle momentum.

The large acceptance of all detectors ($\simeq 2^\circ$ to $\simeq 48^\circ$
horizontally and $\simeq\pm 15^\circ$ vertically),
on both sides of the beam, facilitated the coincident 
measurement of four charged particles, which was crucial for the
reconstruction of many exclusive channels like $ppK^+K^-$~\cite{JIM,JIM2},
$pp\pi^+\pi^-\pi^0$, $pK\Lambda$ and $pK\Sigma$~\cite{DNN}.

The multi-particle trigger~\cite{TRI}, which was based on the multiplicity 
of the hodoscope elements and the scintillating fibers, selected events with at 
least three charged tracks in the final state. The results presented in this 
work are based on $4.2\times10^7$ reconstructed events with
four charged particles (mainly $pp\pi^+\pi^-$) detected.

The exclusive $\eta$ meson production ($pp\rightarrow pp\eta$)
was identified via its dominant decay channel involving charged particles
($\pi^+\pi^-\pi^0$, branching ratio 23.1\%) and 
the reaction $pp\rightarrow pp\eta\prime$ was reconstructed via the 
decay of the $\eta\prime$ meson into $\pi^+ \pi^- \eta$ 
(branching ratio 43.8\%). 
The selection of the exclusive $\eta\prime$ and $\eta$ meson production
is based on two kinematical observables, the 4-particle ($pp\pi^+\pi^-$)
missing mass 
($M_{miss}^{4p}$) and the proton-proton missing mass ($M_{miss}^{pp}$). 
Furthermore the light output from the \v Cerenkov
detectors together with the particle momentum
were used to discriminate between $\pi^+$ mesons and protons
in the final state.

\begin{figure}[tb]
  \begin{center}
   \mbox{\epsfig{figure={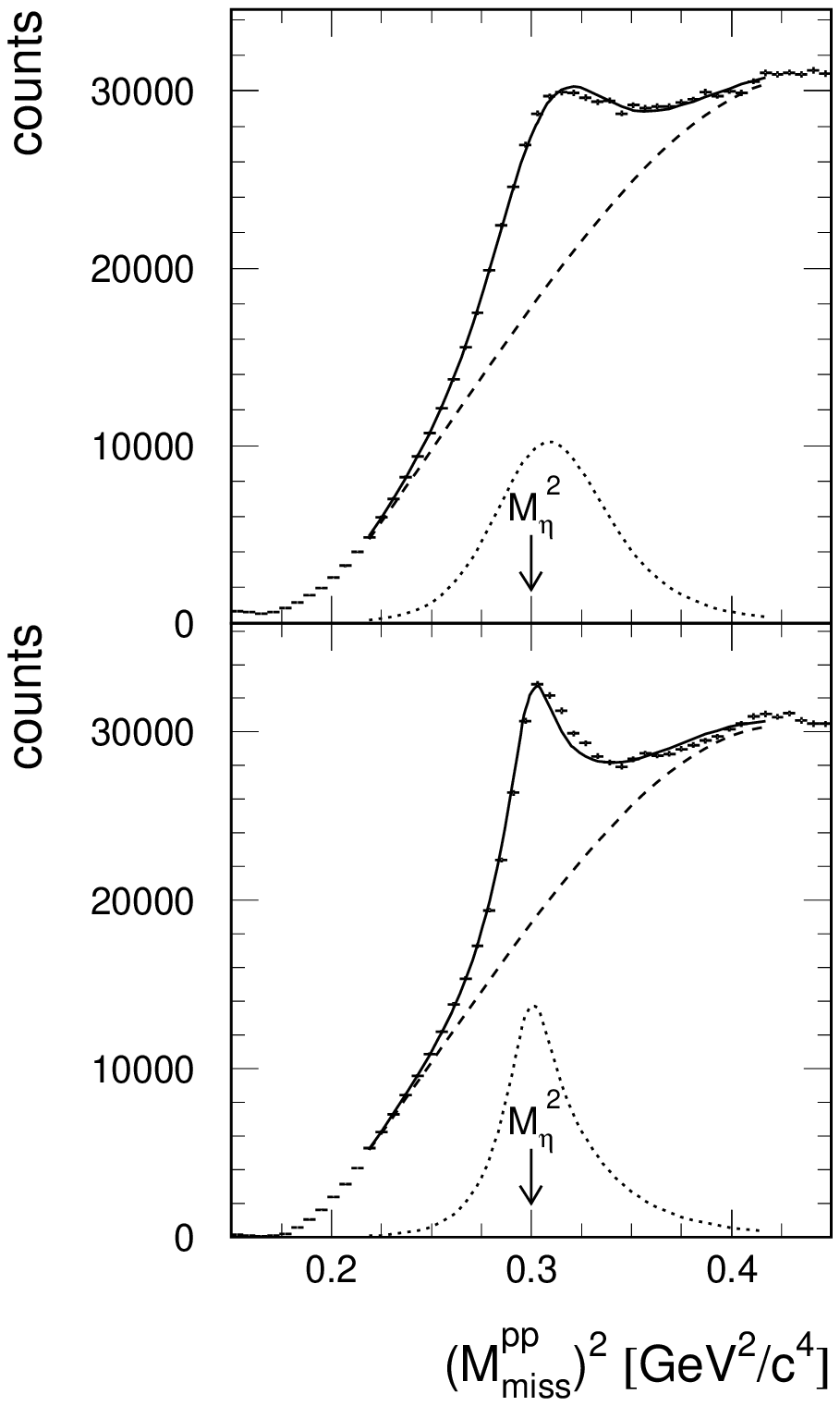},
        width=0.9\linewidth }}
  \end{center}
  \caption{
    Raw spectrum of $(M^{pp}_{miss})^2$ for events with four
    charged particles ($pp\pi^+ \pi^-$) in the final state and 
    a 4-particle missing mass ($M_{miss}^{4p}$) consistent with a 
    missing $\pi^0$ meson. Before (upper frame) and after (lower frame)
    a kinematical refit procedure was applied to the data. 
    The dashed curve represents the background contribution and the solid
    curve shows the sum of the background and of the signal (dotted curve). 
   }
  \label{fig:eta_raw}
\end{figure}

Since neutral pions were not detected 
($\pi^0 \rightarrow \gamma \gamma$, branching ratio 98.8\%),
$pp\eta$ events were selected in which $M_{miss}^{4p}$ was approximately
consistent with a missing $\pi^0$ meson 
($0.005~\mbox{GeV}^2/\mbox{c}^4 < \left(M_{miss}^{4p}\right)^2 < 0.035~\mbox{GeV}^2/\mbox{c}^4 $). 
After imposing this constraint the proton-proton missing mass distribution is shown in
Fig. \ref{fig:eta_raw} (upper frame). A broad signal from  the $\eta$ meson 
is visible near $M_{miss}^{4p}\simeq 0.3~\mbox{GeV}^2/\mbox{c}^4$.

The assumption of a missing $\pi^0$ meson allowed a
constraint ($(M_{miss}^{4p})^2 = M_{\pi^0}^2$) 
to be imposed in order to improve the $(M_{miss}^{pp})^2$
mass resolution by a kinematical refit procedure. 
In this procedure all particle momenta were simultaneous
re-determined under the assumption of a missing $\pi^0$ meson.
After applying the refit procedure, the  
resolution of the $\eta$ meson signal in the $(M_{miss}^{pp})^2$  
distribution is improved by about a factor 2 
(see Fig. \ref{fig:eta_raw}, lower frame).

In both frames the solid curve shows the sum of the 
signal (dotted curve) and the background (dashed curve).
The signal line shape was taken from detailed  
Monte Carlo simulations of the detector performance and the yield was determined by 
scaling the line shape to match the data using a $\chi^2$ minimization
procedure.
The small deviations at $(M_{miss}^{pp})^2 \simeq 0.325~\mbox{GeV}^2/\mbox{c}^4$
result from imperfections of the modeling of the detector response and are
included in the estimation of the systematic errors.

The non-resonant reaction $pp\rightarrow pp\pi^+\pi^-\pi^0$
accounts for most of the background under the $\eta$ meson signal.
Since the exact shape of the background is not quantitatively
known, the background has been parameterized with a $3^{rd}$ order polynomial,
that provides the best $\chi^2$ to the fit.  

The reconstruction of the decay channel
$pp\rightarrow pp\eta\prime \rightarrow pp\pi^+\pi^- \eta$ is analogous
to the $\eta$ meson reconstruction described above, 
since the $\eta$ meson decays mostly by neutral modes
(branching ratio 71.5\%).

\begin{figure}[tb]
  \begin{center}
   \mbox{\epsfig{figure={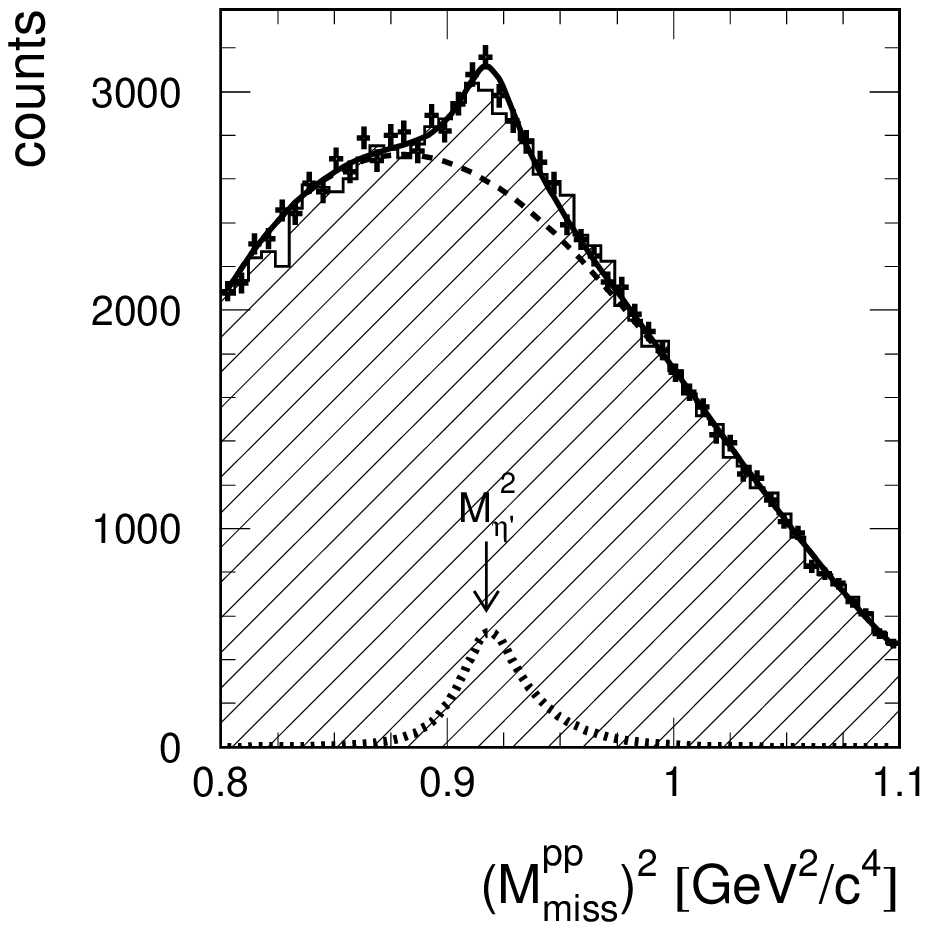},
        width=0.9\linewidth }}
  \end{center}
  \caption{
    Raw spectrum of $(M^{pp}_{miss})^2$ for events with four
    charged particles ($pp\pi^+ \pi^-$) in the final state and 
    a 4-particle missing mass ($M_{miss}^{4p}$) consistent with a 
    missing $\eta$ meson. The data points show the spectrum after a kinematical
    refit procedure was applied. The dashed curve represents the 
    background contribution and the solid curve shows the sum of the
    background and of the signal (dotted curve). In addition the
    hatched spectrum indicates the spectrum before the refit procedure was applied. 
   }
  \label{fig:ep_raw}
\end{figure}

For this decay channel
$M_{miss}^{pp}$ and $M_{miss}^{4p}$ must correspond to a missing 
$\eta\prime$ and a missing $\eta$, respectively. The $(M^{pp}_{miss})^2$ 
distribution for events with a 4-particle missing mass consistent with 
a missing $\eta$ meson 
(i.e. $\left|(M^{4p}_{miss})^2 - M_{\eta}^2\right| < 0.03~\mbox{GeV}^2/\mbox{c}^4 $ )
 is shown in Fig. \ref{fig:ep_raw} before (hatched area) and after (data points)
the kinematical refit procedure
with the constraint $M^{4p}_{miss} = M_{\eta}$.
The spectrum after the refit shows a clear signal from the $\eta\prime$
meson near $M_{\eta\prime}^2\simeq 0.92~\mbox{GeV}^2/\mbox{c}^4$.
In comparison to the reconstruction of the $\eta$ meson the kinematical refit
procedure only improves the resolution of the $\eta\prime$ signal by about
$25\%$, due to the lower laboratory momenta of the outgoing protons.

The dashed curve represents the
background contribution and the solid curve shows the sum of the background
and the signal (dotted curve). The shape of the signal was determined from the  
Monte Carlo simulations and the yield was determined analogously as described
above for the $\eta$ meson yield. The background below the
$\eta\prime$ meson signal originates mainly from non-resonant 
$pp \rightarrow pp \eta\pi^+\pi^-$ production
and reactions such as $pp \rightarrow pp \pi^+\pi^-\pi\pi$ where 
two pions are not detected. Since the exact form is unknown, the shape of the 
background was parameterized as a $4^{th}$ order polynomial, that provides the
best $\chi^2$ to the fit.

The relative acceptance correction of the DISTO spectrometer for the 
$pp\eta\prime$ vs. $pp\eta$ channels has been determined using Monte Carlo
simulations which were analyzed as the measured data. 

The detector acceptance was determined as a 4-dimensional function of 
all relevant degrees of freedom in the $pp\eta\prime$ and
$pp\eta$ final states.

The five degrees of freedom related to the three body decay of the
$\eta\prime$ and the $\eta$ mesons were integrated in the simulations using an
isotropic orientation of the decay plane and the measured matrix elements 
~\cite{MEP,MET}. 

After accounting for azimuthal symmetry in the production reaction, we 
divided the kinematically allowed phase space into
four-dimensional bins and evaluated the efficiency for each bin separately.
This was realized by storing the number of generated and the number of
reconstructed events from the simulations for each phase space bin. The
bin-by-bin ratio provides the efficiency correction, which was
stored in the 4-dimensional acceptance correction matrix.

The acceptance values were typically larger than $1~\%$ in each phase space bin
including branching
ratios, particle decay in flight, tracking efficiency and particle
identification efficiency. 
The approximately $6~\%$ efficiency loss due to the refit procedure was also
accounted for. 
The data were corrected using the appropriate entry from this
4-dimensional acceptance correction matrix, on an event by event basis.
For a detailed discussion
of the relative acceptance correction method see~\cite{JIM3}.

The simulations indicated a very low acceptance of
the apparatus for $\eta$ mesons produced
in the backward hemisphere in the CM (center of mass) rest frame. 
However, since the entrance consists of two identical particles 
a reflection symmetry about $\Theta_{CM}=90^{\circ}$ exists, thus the
backward data are redundant for determining total cross sections.
Therefore, we only analyzed the acceptance-corrected production yield in the 
forward hemisphere, where the acceptance was non-zero in each phase space bin.

It should be noted that
the acceptance correction is essentially independent of the 
event generator used in the simulations due to the complete phase space
coverage of the DISTO spectrometer. 
The generator used for the correction
assumed uniform phase space density for both reactions.
This assumption was verified by using the same correction for 
different phase space populations in the
simulations. The small deviations observed are included in the
estimation of the systematic errors.

Furthermore, the simulations included all important decay
modes for the $\eta$ meson~\cite{PDB} for the acceptance correction 
of the $pp \rightarrow pp\eta\prime \rightarrow pp\pi^+\pi^-\eta$ and the 
$pp \rightarrow pp\eta$ reactions. Hence, background processes such as 
$\eta\prime \rightarrow \pi^+\pi^-\eta \rightarrow \pi^+\pi^-(\pi^+\pi^-\pi^0)$, 
where one or both of the observed pions are from the $\eta$ or events with
more than four charged particles in the acceptance of the detector are
properly accounted for.

After correcting the $\eta\prime$ meson and the $\eta$ meson production yields
for the respective acceptances and branching ratios~\cite{PDB} as described 
above, the measured total cross section ratio
$\mbox{R}=\sigma_{pp\rightarrow pp\eta\prime} / \sigma_{pp\rightarrow pp\eta}$ is
determined to be
$(0.83\pm0.11^{+0.23}_{-0.18})\times 10^{-2}$. Where the first error is
statistical and the second error range is due to systematic uncertainties.
Because both meson channels have been reconstructed in events with the same
four-charged-particle final state and measured simultaneously within the same
experiment, many systematic uncertainties cancel when considering the
production ratio.  As a result, the systematic error is dominated by the
background subtraction and the relative acceptance correction.

\begin{figure}[tb]
  \begin{center}
    \mbox{\epsfig{figure={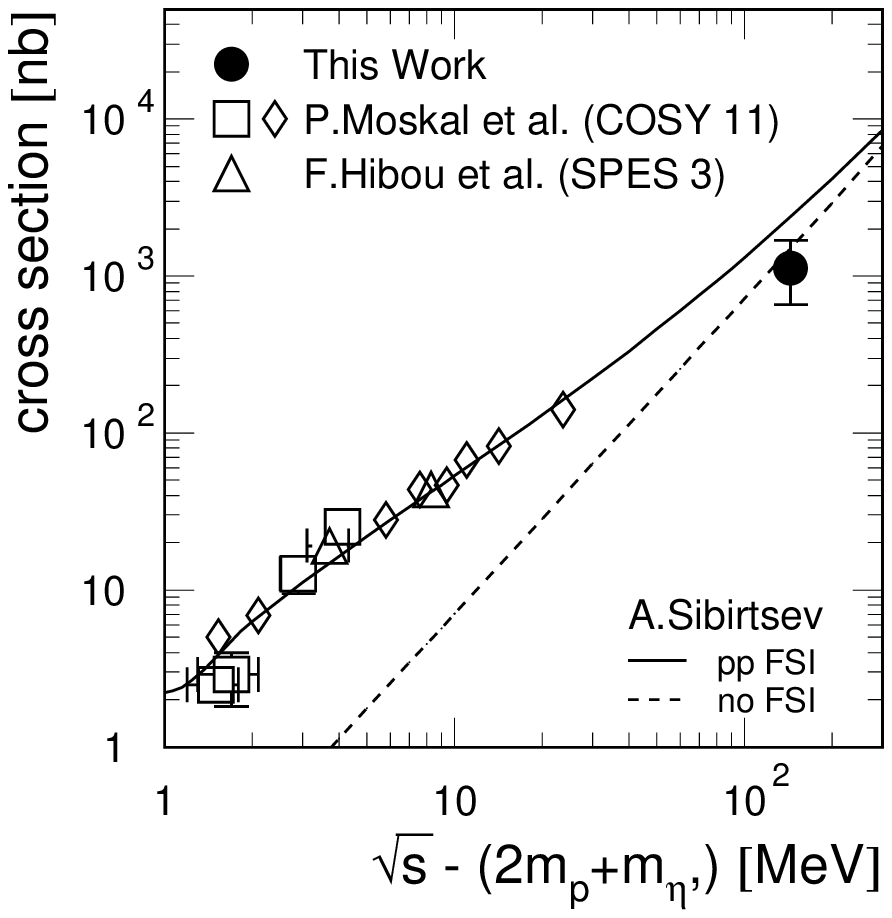},width=0.9\linewidth }}
  \end{center} 
  \caption{
      Total cross section for the $pp \rightarrow pp\eta\prime$
      reaction as a function of the available energy
      above the $\eta\prime$ production threshold.
      Shown is the value measured in this work (full circle) together
      with other data and model calculations described in the text.
    }
  \label{fig:ep_wq}
\end{figure}

The total cross section for the $pp \rightarrow pp\eta$ reaction
is known over a wide energy range \cite{ETA-D} above and below 
the beam momentum of this measurement.
Interpolation of the existing data leads to a   
cross section of $\sigma^{exp}_{pp \rightarrow pp\eta}=135\pm35~\mu b$
at $p_{beam}=3.67~\mbox{GeV}/\mbox{c}$ in good agreement of the value
$\sigma^{model}_{pp \rightarrow pp\eta}\simeq 120~\mu b$ calculated by
Vetter et al.~\cite{VET} using a meson exchange model.
By multiplying $\mbox{R}$ by $\sigma^{exp}_{pp \rightarrow pp\eta}$
we obtain the total
cross section $\sigma_{pp \rightarrow pp\eta\prime} =
1.12\pm0.15^{+0.42}_{-0.31}~\mu b$.
The systematic error in $\sigma^{exp}_{pp \rightarrow pp\eta}$ is
geometrically added with the systematic error in  the $pp\rightarrow
pp\eta\prime / pp\rightarrow pp\eta$ ratio.
This result is shown in Fig.~\ref{fig:ep_wq} (filled circle)
together with other data closer to the production threshold
\cite{MOS,HIB,MOS2} and model calculations (solid and dashed curves)~\cite{SIB}.

The calculation from Sibirtsev et al.~\cite{SIB} represents a 
one-pion-exchange model including the $pp$ FSI. 
The solid line denotes the full calculation and
the dashed line corresponds to the same calculation excluding the 
FSI. The full calculation reproduces the near threshold 
data well, but predicts a cross section of about $2.3~\mu b$ at our 
energy, which is significantly above our measurement.

\begin{figure}[tb]
  \begin{center}
    \mbox{\epsfig{figure={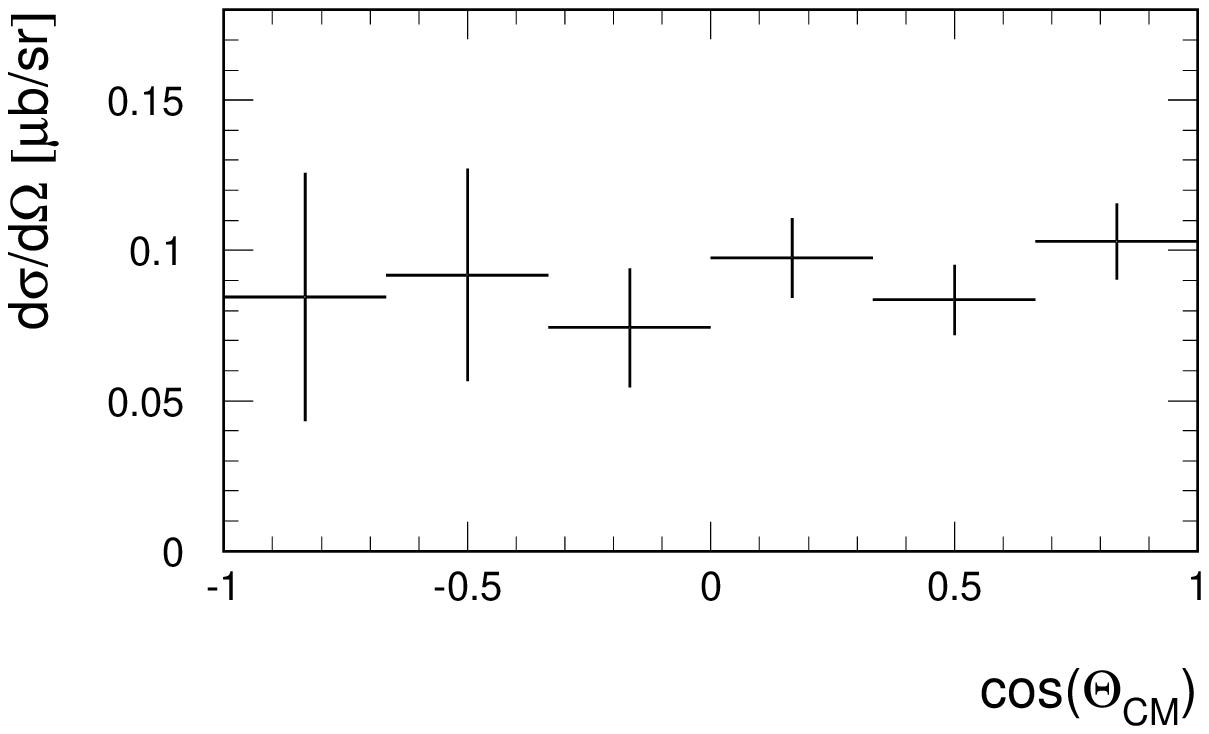},width=0.9\linewidth }}
  \end{center}
  \caption{
    Differential cross section for the $pp \rightarrow pp\eta\prime$ reaction
    as a function of the CM polar angle of the $\eta\prime$ meson 
   }
  \label{fig:ep_ct}
\end{figure}

\begin{figure}[tb]
  \begin{center}
   \mbox{\epsfig{figure={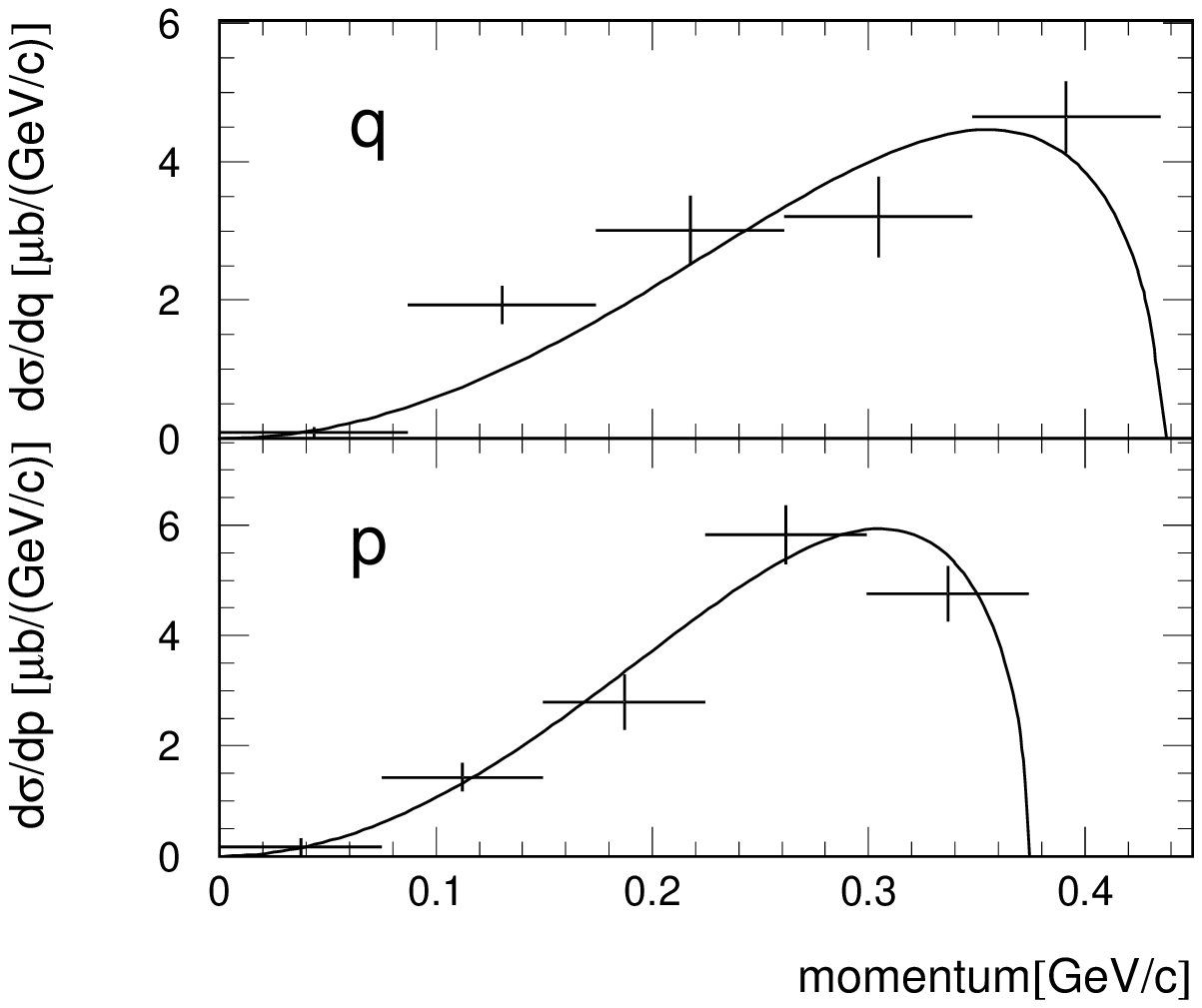},width=0.9\linewidth,clip=}}
  \end{center}
  \caption{
    Differential cross section for the $pp \rightarrow pp\eta\prime$ reaction
    as a function of the $\eta\prime$ meson momentum in the CM frame
    ($q$, upper frame) and the proton momentum in the $pp$-rest frame
    ($p$, lower frame).
    The curves denote the behavior of the three body phase space when the two
    protons are in a relative s-wave state and the $\eta\prime$ meson is in a 
    s-wave state relative to the protons.
   }
  \label{fig:ep_qp}
\end{figure}

The distribution of the differential cross section as a function of  
the CM polar angle of the $\eta\prime$ meson ($\cos(\Theta_{CM})$)
is shown in Fig.~\ref{fig:ep_ct}. The distribution displays no significant 
deviations from isotropy, indicating that the $\eta\prime$ meson is 
predominantly in a s-wave state relative to the two protons.
The differential cross section distributions were determined by producing the
corresponding $(M_{miss}^{pp})^2$ spectra for each kinematical bin.
Each spectrum was fitted 
analogously as described above to determine the yield of the signal 
for each bin. 
The statistical uncertainty of the yield for each bin are
determined by the fitting procedure and are shown as 
vertical error bars in the differential distributions (see
Fig.~\ref{fig:ep_ct} and Fig.~\ref{fig:ep_qp}).
The signal line shape was calculated for each spectrum individually from the
Monte Carlo simulations and the background was also allowed to vary. 
 
The relative partial wave contributions in the three particle final state
$pp\eta\prime$ can also be determined from the momentum distributions $q$ and
$p$, where $q$ is the CM momentum of the $\eta\prime$ meson 
and $p$ is the momentum of a proton in the
proton-proton rest frame.

Then, the total cross section is given as the sum of the individual partial
wave contributions \cite{MEY}:
\begin{equation}
\label{eq:sigma}
\sigma \sim \sum_{l_1,l_2} \int  |M_{l_1,l_2}|^2 d\rho_{l_1l_2}.
\end{equation}
here the sum extends over the angular momenta $l_1,l_2$ and 
the transition amplitude for the given exit channel ($M_{l_1,l_2}$)
, where $l_1$ is
the orbital angular momentum of the two protons relative to each other and
$l_2$ is the orbital angular momentum of the $\eta\prime$ meson relative to
the proton-proton system. 

The corresponding phase space element $d\rho_{l_1l_2}$ is determined by 
\begin{equation}
\label{eq:rho}
d\rho_{l_1l_2} \sim p^{2l_1+1}q^{2l_2+2}dq,
\end{equation}

If we assume the transition amplitude $M_{l_1,l_2}$ to be almost constant
over the available phase space then the differential cross section
as a function of the particle momenta are given by the variation of the phase 
space volume.

The measured differential cross sections are plotted in Fig.~\ref{fig:ep_qp}.
as a function of $q$ (upper frame) and $p$ (lower frame). 
The corresponding curves, that reproduce the data quite well, have been
obtained from equation~\ref{eq:rho}, assuming $l_1=l_2=0$. The normalization
was obtained by a simultaneous $\chi^2$ minimization to both differential cross section 
distributions. The introduction of higher values of $l$ does 
not improve the $\chi^2$ fit to the data. The result is consistent
with a dominant Ss-wave production of the $\eta\prime$ meson, where S denotes
$l_1=0$ and s $l_2=0$, respectively.

In conclusion, the production of the pseudoscalar $\eta\prime$
meson has been studied in the $pp$ reaction at $p_{beam} = 3.67~\mbox{GeV/c}$.
The $\eta\prime$ meson has been reconstructed by measuring
its charged decay products.
The $\eta\prime /\eta$ cross section ratio has been determined and the 
extracted $\eta\prime$ cross section has been compared to data very close to
threshold and an one-pion exchange model including the $pp$ FSI. While the
model describes the data close to threshold very well it overestimates our data
point by about $100~\%$.

Furthermore, the differential cross sections indicates that the $\eta\prime$ meson is
predominantly produced in a s-wave state
for the two protons relative to each other and the $\eta\prime$ meson relative to
the proton-proton system.   

Calculations up to $100~\mbox{MeV}$ above the $\eta\prime$ production threshold, 
in the framework of a relativistic
meson-exchange model~\cite{NAG} should be extended to higher energies. 
Comparison with our results should help to learn more about the
different $\eta\prime$ meson production mechanisms, the potential
influence of a $\eta\prime p$ FSI~\cite{BAR} and the magnitude of the coupling
constant $g_{\eta\prime pp}$.

In this context a comparison with a recent publication on the production of
$\pi^0$, $\eta$ and $\eta\prime$ mesons in proton-proton collisions~\cite{MOS3} 
should be very helpful. The novel approach therein 
factored out the $pp$ FSI and the initial-state proton-proton interaction (ISI)   
from the total cross section to determine the phase space dependence of the
total production amplitude for $\pi^0$ ($|A_0^{\pi}|$), $\eta$ ($|A_0^{\eta}|$)
and $\eta\prime$ mesons ($|A_0^{\eta\prime}|$).

Our measurement will assist to evaluate different models used for the
parameterization of the $pp$ FSI which is essential for the determination
of the absolute strength of $|A_0^{\eta\prime}|^2$ and hence of $g_{\eta\prime pp}$.

This paper is dedicated to the memory of Igor Falomkin.

We acknowledge the work provided by the SATURNE II accelerator staff and
technical support groups in delivering an excellent proton beam 
and assisting this experimental program. 

This work has been supported in part by:
CNRS-IN2P3, CEA-DSM, NSF, INFN, KBN (2 P03B 117 10 and 2 P03B 115 15)
and GSI.


\begin{thebibliography}{99} 

\bibitem{WEI}
S.~Weinberg, Phys.~Rev.~D~11~(1975)~3583. 

\bibitem{WIT}
E.~Witten, Nucl.~Phys.~B~156 (1979)~269.

\bibitem{VEN}
G.~Veneziano, Nucl.~Phys.~B~159 (1979)~213.

\bibitem{HOF}
G.~t'Hooft, Phys.~Rev.~Lett.~37~(1976)~8;
G.~t'Hooft, Phys.~Rev.~D~14~(1976)~3432.

\bibitem{GLU}
A.De~Rejula, H.~Georgi and S.L.~Glashow, Phys.Rev.~D~12~(1975)~147.

\bibitem{CLEO-D}
B.H.~Behrens~et~al., Phys.~Rev.~Lett.~80~(1998)~3710;
T.E.~Browder~et~al., Phys.~Rev.~Lett.~81~(1998)~1786.

\bibitem{CLEO-T}
D.~Atwood and A.~Somi, Phys.~Rev.~Lett.~79~(1997)~5206.

\bibitem{EMC}
J.~Ashman~et~al., Phys.~Lett.~B~206~(1988)~364.

\bibitem{HAT}
T.~Hatsuda, Nucl.~Phys.~B~329~(1990)~376.

\bibitem{SHO}
G.M.~Shore and G.~Veneziano, Phys.~Lett.~B~244~(1990)~75.

\bibitem{MOS}
P.~Moskal~et~al. Phys.~Rev.~Lett~80~(1998)~3202.

\bibitem{HIB}
F.~Hibou~et~al. Phys.~Lett.~B~438~(1998)~41.
 
\bibitem{MOS2}
P.~Moskal~et~al. Phys.~Lett.~B~474~(2000)~416.

\bibitem{VET}
T.~Vetter et al., Phys.~Lett.~B~263~(1991)~153.

\bibitem{MEI}
V.~Bernard, N.~Kaiser and U.-G.~Mei\ss ner, Eur.~Phys.~J.~A~4~(1999)~259.

\bibitem{SIB}
A.~Sibirtsev and W.~Cassing, Eur.~Phys.~J.~A~2~(1998)~333.

\bibitem{ETAFSI}
G.~F\"aldt and C.~Wilkin, Z.~Phys.~A~357~(1997)~241.

\bibitem{NIM}
F.~Balestra~et~al., Nucl.~Instr.~Meth.~A~426~(1999)~385.  

\bibitem{JIM}
F.~Balestra et~al., Phys.~Rev.~Lett.~81~(1998)~4572.

\bibitem{JIM2}
F.~Balestra~et~al., Phys.~Lett.~B468~(1999)~385.

\bibitem{DNN}
F.~Balestra~et~al., Phys.~Rev.~Lett.~83~(1999)~1534.

\bibitem{TRI}
F.~Balestra~et~al., IEEE Trans.~Nucl.~SCI.~45~(1998)~817.

\bibitem{MEP}
G.R.~Kalbfleisch, Phys.~Rev.~10~(1974)~916.

\bibitem{MET}
C.~Amsler~et~al., Phys.~Lett.~B~346~(1995)~203.

\bibitem{JIM3}
F.~Balestra~et~al., submitted to Phys.~Rev.~C.

\bibitem{PDB}
C.~Caso~et~al., Eur.~Phys.~J.~C~3~(1998)~1. 

\bibitem{ETA-D}
 J. Smyrski~et~al., Phys.~Lett.~B~474~(2000)~182;
 H. Cal\'an~et~al., Phys.~Lett.~B~366~(1996)~39;
 A. M. Bergdolt~et~al., Phys.~Rev.~D~48~(1993)~R2969;
 E. Chiavassa~et~al., Phys.~Lett.~B~322~(1994), 270 (1994);
 E. Pickup~et~al, Phys.~Rev.~Lett.~8~(1962)~329;
 L. Bodini~et~al., Nuov.~Cim.~58~A~(1968)~475;
 A. P. Colleraine and U. Nauenberg, Phys.~Rev.~161~(1967)~1387;
 G. Alexander~et~al., Phys.~Rev.~154~(1967)~1284;
 C. Caso~et~al., Nuov~Cim.~55~A~(1968)~66;
 E. Colton and E. Gellert, Phys.~Rev.~D~1~(1970)~1979;
 G. Yekutie~et~al., Nucl.~Phys.~B~18~(1970)~301;
 S.P. Almeida~et~al., Phys.~Rev.~174~(1968)~1638;
 J. Le Guyader~et~al., Nucl.~Phys.~B~35~(1971)~573.

\bibitem{MEY}
 R.G.~Newton, Scattering Theory of Waves and Particles, Springer Verlag, New York.
 H.O.~Meyer, Particles and Fields Series 41, AIP Conference Proceedings, New York.

\bibitem{NAG}
K.~Nakayama~et~al., Phys.~Rev.~C~61~(2000)~024001. 

\bibitem{BAR}
V.~Baru~et~al., Eur.~Phys.~J.~A~6~(1999)~445.

\bibitem{MOS3}
P.~Moskal~et~al. Phys.~Lett.~B~482~(2000)~356.

\end{thebibliography}
\end{document}